\documentclass{article}
\usepackage{graphicx} 

\counterwithin*{section}{part}

\usepackage[utf8]{inputenc}
\usepackage{amsmath,amssymb,amsfonts}
\usepackage{mathtools}
\usepackage{boldmath}
\usepackage{xcolor}

\usepackage{booktabs}

\usepackage{siunitx}
\usepackage{graphicx}
\usepackage{caption}
\usepackage{subcaption}
\usepackage[hidelinks]{hyperref}

\newcommand\nnfootnote[1]{%
  \begin{NoHyper}
  \renewcommand\thefootnote{}\footnote{#1}%
  \addtocounter{footnote}{-1}%
  \end{NoHyper}
}

\DeclareMathOperator*{\argmax}{argmax}

\usepackage{flushend}
\usepackage{tikz}
\usetikzlibrary{shapes.geometric}
\usepackage[RPvoltages]{circuitikz}
\usepackage{tikzset} 

\RequirePackage[T1]{fontenc}
\usepackage{comment}

\renewcommand{\L}{\mathcal{L}}

\usepackage{biblatex}
\usepackage[nameinlink,noabbrev,capitalise]{cleveref}

\title{Automatic Control of\\Reactive Brain Computer Interfaces}

\author{Pex Tufvesson$^1$\thanks{Ericsson Research, Lund, Sweden. Department of Automatic Control, Lund University, Sweden. Corresponding Author: pex.tufvesson@control.lth.se or pex.tufvesson@ericsson.com} \and Frida Heskebeck$^1$\thanks{Department of Automatic Control, Lund University, Sweden. frida.heskebeck@control.lth.se}}
\date{October 2023}

\begin{document}
\sloppy
\maketitle
\emph{Contribution to the field}. This journal paper presents an approach toward calibration-free reactive brain computer interfaces. A grand challenge for current BCI research is establishing efficient methods for reliable online classification of neural activity. Our approach is a step towards ready-to-use brain computer interfaces with the potential of expanding the boundaries of BCI appliances and research.

\addtocounter{footnote}{+1}

\nnfootnote{$^1$The authors contributed equally and share first authorship.}

\section*{Abstract}

This article discusses practical and theoretical aspects of real-time brain computer interface control methods based on Bayesian statistics.
We investigate and improve the performance of automatic control and feedback algorithms of a reactive brain computer interface based on a visual oddball paradigm for faster statistical convergence. We introduce transfer learning using Gaussian mixture models, enabling a ready-to-use setup.

\section*{Keywords}

Brain Computer Interface, Automatic Control, Gaussian Mixture Model, Bayesian Statistics, Transfer Learning, Monte Carlo Simulation

\begin{figure}[h!]
\begin{center}
\includegraphics[width=8.4cm]{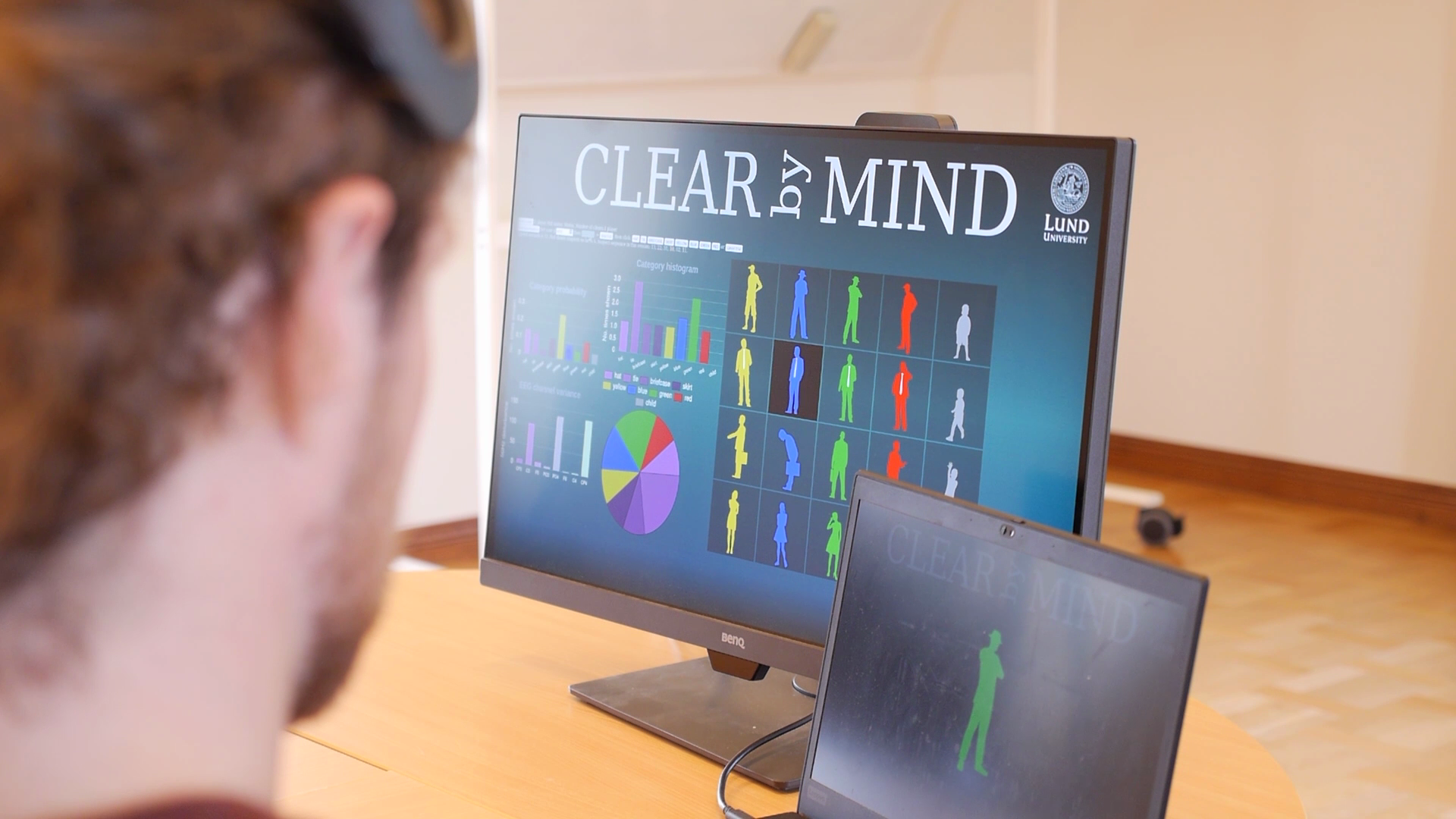}
\caption{Visual color stimuli used in the Clear by Mind brain game from the BCI-HIL research framework published by \cite{gemborn:23}. This is an example of a reactive BCI answering the question \emph{What color am I thinking of?}}
\label{fig:cbm}
\end{center}
\end{figure}

\section{Introduction}

One of the purposes of a Brain Computer Interface (BCI) is to decode the user's brain activity to understand their intentions. The book chapter by \textcite{nam_braincomputer_2018-ch1} gives an excellent introduction to BCIs. In a so-called reactive BCI, the user's brain activity in response to given stimuli is measured and analyzed. The stimuli could, for example, be a sequence of images of colored silhouettes, where one color is the \emph{target stimulus}, and all other stimuli are \emph{non-target stimuli}. The BCI aims to identify the user's selected target stimulus based on the measured brain activity. It is desirable to identify the target stimulus as fast as possible. To do that, the sequence of stimuli should be chosen to maximize knowledge gain. Adaptive sequence selection for so-called P300 experiments has been attempted before, for example, in \cite{ma_adaptive_2021} where they framed the problem as a multi-armed bandit problem or in \cite{grizou_calibration-free_2014} where they used hypothesis testing. In this paper, we use Bayesian statistics based on Gaussian mixture models. 

Most BCIs measure brain activity by electrodes placed on the outside of the head. Today, the applications are mainly medical, such as helping paralyzed individuals regain some form of motor function. There have been instances where such individuals have used BCIs to type \cite{rezeika_braincomputer_2018}, control robotic arms \cite{vilela_chapter_2020}, control wheelchairs \cite{zhang_control_2015}, or even walk using exoskeletons \cite{colucci_braincomputer_2022}. Some slightly more futuristic applications are brain-to-text \cite{willett_high-performance_2021}, artificial retina \cite{muratore_artificial_2020}, and passthoughts \cite{merrill_one-step_2019}.
Many people have heard about Neuralink \cite{neuralink_neuralink_2023}, now recruiting patients for their six-year-study on EEG implants for persons with quadriplegia due to cervical spinal cord injury or amyotrophic lateral sclerosis (ALS). The commercial market for BCIs is expanding, with several companies offering consumer-grade devices for meditation \cite{acabchuk_measuring_2021}, mental training \cite{mitsea_brain-computer_2023}, or gaming which allows game interaction using thought or mood-based inputs \cite{varjo}. While not as advanced as clinical-grade equipment, these devices are making BCI technology more accessible to the general public.

\cref{sec:EEG_to_GMM} in this paper presents how the measured brain signals are mapped to a real number that we want to estimate the distribution of using a Gaussian mixture model (GMM). The section also presents the experimental setup and the used dataset. In \cref{sec:stimuli_control}, the GMMs are used for optimal sequence selection to minimize the number of shown stimuli needed to identify the target stimulus. \cref{sec:transfer_learning} highlights the difference in measured data between sessions and suggests an approach for transfer learning based on the GMMs. Sections \ref{sec:Results1}, \ref{sec:Results2} and \ref{sec:Results3} present the results from the paper, \cref{sec:Discussion} is our discussion, and \cref{sec:conclusions} concludes the paper.

\subsection{The human brain response}

Hans Berger recorded the first human electroencephalograms (EEGs) and published many papers on the topic. His first paper on human EEG was published in 1929 \cite{berger_uber_1929}. EEG records the voltage difference between electrodes placed on the user's scalp. The recorded data, $\BX_{EEG} \in \mathbb{R}^{m,n}$, is a time series data where $m$ corresponds to the number of electrodes, also called channels, and $n$ to the number of samples. The data is often sampled at about 500~Hz \cite{lopes_niedermeyer_2012}, and the most studied EEG rhythms like delta, theta, alpha, beta, and gamma waves reside below 60~Hz. The electrodes are positioned in a so-called 10-20 system on a cap, and conductive gel is used to improve the connection between the electrodes and the skin. These are called wet electrodes, while dry electrodes also exist which don't require any conductive gel but generally have a looser connection to the skin with higher impedance and are more prone to artifacts than the wet electrodes \cite{afif_comparison_2020}. In a typical BCI experiment, the user is stationary in front of a screen where instructions are given to the user.

Different types of brain activity, so-called paradigms, can be used as ``control signals'' in BCIs. The P300 response is a so-called \emph{event-related potential}, which is one of the most common type of BCI paradigm \cite{abiri_comprehensive_2019}. When a user is given a stimulus, for example, shown an image, different waveforms can be registered in the measured brain signals in response to the stimuli. The waveform at around 300~ms after stimuli onset, the moment when the stimulus is shown, is related to the information value of the stimulus. An increased level of attention to the stimulus by the user corresponds to a heightened amplitude in the P300 wave \cite{sur_event-related_2009}. The P300 signal was discovered in the 1960's \cite{sutton_evoked-potential_1965} and was used in a so-called P300-speller about 20 years later \cite{farwell_talking_1988}. In the original P300-speller, the alphabet is arranged in a grid, and the rows/columns of the grid are flashed alternately. The user focuses on the letter they want to print, and a P300 signal can be detected when that specific letter is flashed.

\begin{figure}[h!]
\begin{center}
\begin{subfigure}[t]{0.49\linewidth}
    \includegraphics[width=\linewidth]{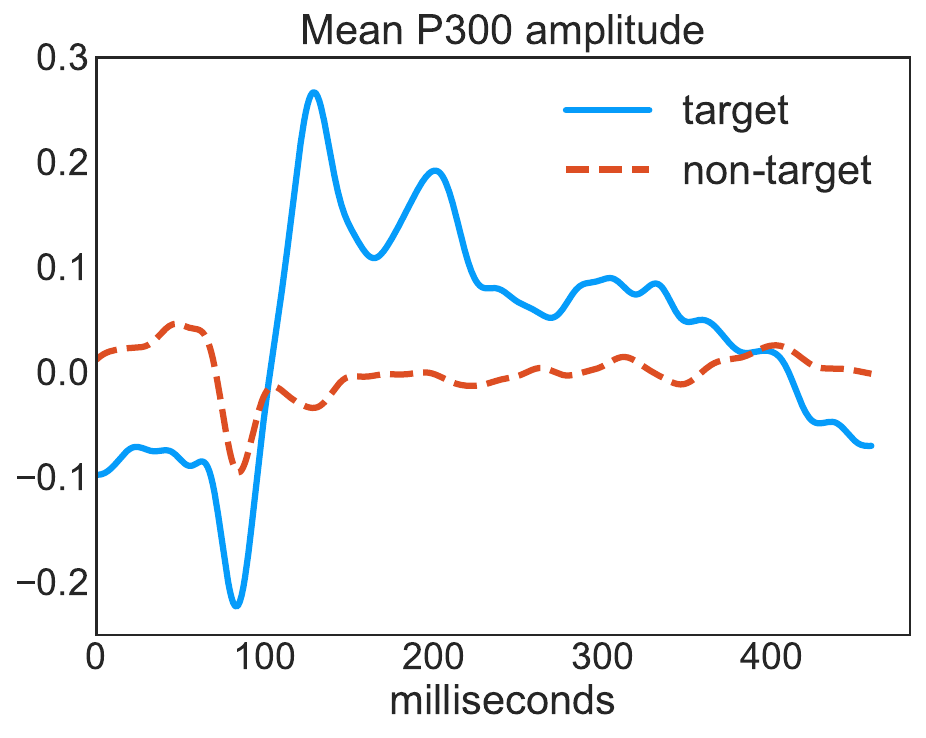}
    \caption{Average of P300 signals.}
    \label{subfig:P300_mean}
\end{subfigure}
\begin{subfigure}[t]{0.49\linewidth}
    \includegraphics[width=\linewidth]{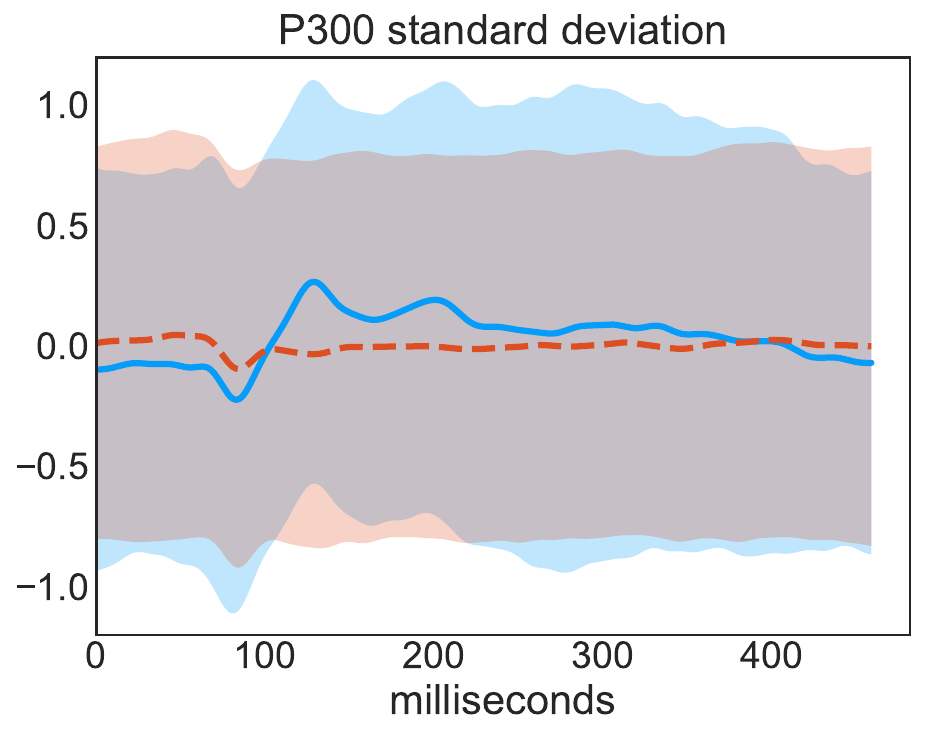}
    \caption{Average of P300 signals with standard deviation.}
    \label{subfig:P300_std}
\end{subfigure}
\caption{P300 response for target (blue) and non-target (red) data. 
The P300 response in (a) seems easy to classify since the target and non-target amplitudes differ. However, the reason for using multiple stimuli and response iterations is evident when showing the standard deviation, as seen in (b).}
\label{fig:mean_and_std}
\end{center}
\end{figure}

The P300 response works well on average, as seen in \cref{subfig:P300_mean}. But as \cref{subfig:P300_std} shows, the standard deviation is huge. Thus, a single measurement of the P300 signal is not enough to detect if the shown stimulus was attended to, but multiple measurements are needed. In addition, more EEG channels and more complex signal analysis could also help.

One key assumption taken when doing statistical analysis on any signal, which in our case means feature map values gotten from EEG P300 signals, is the Gaussian distribution. Ideally, we also require successive epochs to be independent and identically distributed \cite{luck_introduction_2014}. Depending on the experimental setup, this can be more or less true. EEG is prone to slow drifts and artifacts that span multiple epochs. Also, the experiment may have too short duration of the stimuli, together with too short pauses between stimuli, making the P300 response contain superimposed responses from previous epochs. Also, note that the removal of visual stimuli is a stimulus itself \cite{woodman_brief_2010} - in a reactive brain computer interface paradigm, the human brain will react to visual stimuli being removed. On the other hand, using simple stimuli at a too slow rate will make the human user feel bored, which leads to lower performance \cite{jin_improved_2017}. If the odd-one-out stimulus we want to detect is shown too frequently, the P300 response gets weaker \cite{duncan-johnson_quantifying_1977}.

The human brain is a massively parallel non-linear computational device, and EEG is a blunt tool to measure its operation. Besides the difficult task of supervising the operation of 86 thousand million neurons using fewer than 100 electrical measuring points, the human user will be distracted by other mental images popping up, the mood and level of excitement, and bodily functions such as the need for eye blinks and voluntary and involuntary muscle movements. The experiment design can reduce some of these artifacts, like adding pauses for eye-blinks, but not all of them \cite{luck_introduction_2014}.

\section{From EEG epochs to Gaussian mixture models} \label{sec:EEG_to_GMM}
This section describes the theory, materials and methods, calculations, and results for generating Gaussian mixture models from EEG data. 

\subsection{Theory}

\subsubsection{EEG Feature maps}\label{sec:EEG_feature_maps}
The measured multi-channel EEG time-series data $\BX_{EEG}$ can be translated into a real number $y \in \mathbb{R}$ through a \emph{feature map} $\Phi: \BX_{EEG} \mapsto y$. We denote $f_0$ as the statistical distribution of $y$ for non-target data. We denote $f_1$ as the statistical distribution of $y$ for target data. Both are \emph{probability density functions} (PDFs). 
The mapping can be done in several ways, and any function that maps EEG data to a real number could serve as a feature map $\Phi$. Some examples are:
\begin{itemize}
    \item \emph{Peak amplitude}: Uses the fact that the target stimuli's P300 waveforms generally have higher amplitudes than the non-target stimuli's P300 waveforms \cite{nam_gentle_2018}, as seen in \cref{subfig:P300_mean}.
    \item \emph{Area under the curve} (\textbf{AUC}): Uses the fact that the target stimuli's P300 waveforms generally have both higher and lower amplitudes than the non-target stimuli's P300 waveforms. AUC integrates the absolute value of the EEG signal over the epoch, which separates target epochs from non-target epochs \cite{mcdowell_aging_2003}.
    \item \emph{Logistic regression} (\textbf{LR}): Calculates a weighted sum of the inputs, EEG data in this case, and outputs the logistic, which is a number between 0 and 1 corresponding to the probability of the target class \cite{geron_training_2019}.
    \item \emph{Linear Discriminant Analysis} (\textbf{LDA}): Finds a separating hyperplane between the classes with regard to the classes' distributions \cite{nam_gentle_2018}.
    \item \emph{Support Vector Machines} (\textbf{SVM}): Finds a separating ``road'' between the classes where the road is as wide as possible \cite{geron_support_2019}.
    \item \emph{xDAWN + LDA} (\textbf{xDAWN}): This is a spatial filtering method used on EEG data to enhance the signal-to-noise ratio \cite{rivet:09}.
    \item \emph{ERP covariance matrices + Logistic regression on Tangent space to Riemannian manifold} (\textbf{ERPcov TS LR}): The EEG covariance matrices live on the Riemannian manifold of positive semidefinite matrices. A feature map in Euclidian space like LR can be applied in the Euclidian tangent space to the Riemannian manifold \cite{barachant_multiclass_2012}.
    \item \emph{ERP Covariance matrices + Minimum Distance to Mean} (\textbf{ERPcov MDM}): This is a feature map adjusted to the Riemannian geometry, in this case, minimum distance to mean. The feature map is applied to the data on the Riemannian manifold \cite{barachant_multiclass_2012}.
\end{itemize}
Due to the noisy nature of EEG signals, the separation of target and non-target data will not be perfect, and the PDFs of non-targets $f_0$ and targets $f_1$ will overlap, regardless of the algorithm chosen to separate them. The feature map algorithms in the above list grow in complexity the further down in the list we come. The two uppermost feature maps need no prior training data and can be used directly. The others need \emph{training data}, \emph{validation data} and can then be used on \emph{test data} \cite{hastie_elements_2009}. The training data is used for fitting the feature maps, validation data is used for hyperparameter selection of the feature maps, and test data transformed with the trained feature maps is used for creating the PDFs. 

Statistical analysis becomes easier with Gaussian independent and identically distributed signals. To be able to handle non-Gaussian probability density functions, we will approximate them using multiple Gaussian distributions, \emph{Gaussian mixture models} (GMMs). The GMM comprises several Gaussian distributions, each parameterized by a mean and covariance matrix \cite{bishop_pattern_2006}. In the context of a two-dimensional GMM for P300 signals, each Gaussian represents a cluster in the feature space. Mathematically, the GMM can be represented as:
\[
p(\mathbf{x}) = \sum_{i=1}^{K} \pi_i \mathcal{N}(\mathbf{x}|\mathbf{\mu}_i, \mathbf{\Sigma}_i)
\]
where \(K\) is the number of Gaussians in the mixture, \(\pi_i\) denotes the mixing coefficient of the \(i^{th}\) Gaussian ($\sum_{i=1}^K\pi_i=1$), \(\mathbf{\mu}_i\) is its mean, and \(\mathbf{\Sigma}_i\) is the covariance matrix. We use the expectation-maximization (EM) algorithm \cite{dempster_em_1977} to fit a two-dimensional GMM with three Gaussian components for the target and non-target data respectively \cite{do_what_2008, ng_em_2012}. Fitted GMMs with corresponding histograms for each feature map can be seen in \cref{fig:eeg_to_gmm} and \cref{fig:pdfs}.

\subsubsection{Scoring a feature map}\label{sec:scoring}
As discussed above, there are many candidates for feature maps, and we have combined them two and two in \cref{fig:pdfs}. We would benefit from including every single feature having a higher than chance accuracy, however, ranking features will help us select the better ones. We perform the ranking by computing the Kullback-Leibler divergence between the target and non-target GMMs, where a higher number means that the GMMs from target and non-target data have a higher degree of separation \cite{kullback_information_1951}. Combining additional feature maps into the GMMs would increase the computation complexity but could improve the separation. The theory and principles for two-dimensional GMMs and higher-order GMMs are similar \cite{bishop_pattern_2006}, and for the sake of visualizations printed on paper, we use two-dimensional GMMs. 

For two 1D Gaussian distributions, each consisting of a single mean and variance, the KL divergence is
\[ D_{KL}(P || Q) = \frac{1}{2} \left( \log \left( \frac{\sigma_Q^2}{\sigma_P^2} \right) + \frac{\sigma_P^2 + (\mu_P - \mu_Q)^2}{\sigma_Q^2} - 1 \right), \]
where \(P\) is a Gaussian with mean \(\mu_P\) and variance \(\sigma_P^2\), and \(Q\) is a Gaussian with mean \(\mu_Q\) and variance \(\sigma_Q^2\).

Comparing two 2D GMMs is a non-trivial task since GMMs are complex probabilistic models. It is worth noting that computing the KL divergence directly between two GMMs has no closed-form solution and can be computationally intensive \cite{hershey_approximating_2007-1}. Direct computation becomes intractable, and for two GMMs, say \(P\) and \(Q\), each with weights \(\pi\) and means \(\mu\) and variances \(\sigma^2\) we have
\[ P(x) = \sum_{i=1}^{M} \pi_i \mathcal{N}(x; \mu_i^P, \sigma_i^P) \]
\[ Q(x) = \sum_{j=1}^{N} \pi_j \mathcal{N}(x; \mu_j^Q, \sigma_j^Q), \]
and the KL divergence between \(P\) and \(Q\) is
\[ D_{KL}(P || Q) = \int P(x) \log \left( \frac{P(x)}{Q(x)} \right) dx .\]

In practice, approximations or sampling-based methods are used to compute the KL divergence for GMMs \cite{hershey_approximating_2007-1}. For instance, using Monte Carlo sampling, we can estimate KL divergence by sampling points \(x_1, x_2, ... x_n\) from \(P\) and then approximating the integral above as
\[ D_{KL}(P || Q) \approx \frac{1}{n} \sum_{i=1}^{n} \log \left( \frac{P(x_i)}{Q(x_i)} \right) .\]
Monte Carlo sampling provides an approximation to the true KL divergence, and its accuracy can be adjusted by varying the number of samples $n$. Increasing $n$ generally improves accuracy but also increases computation time.

Some other distribution distance metrics that could be used include Bhattacharyya distance \cite{bhattacharyya1943measure}, Jeffrey divergence and Jensen-Shannon divergence \cite{cover_elements_2005}, earth mover's (Wasserstein) distance \cite{rubner_earth_2000}, iterative closest point \cite{besl_method_1992}, and cross likelihood \cite{cover_elements_2005}. The Kullback-Leibler divergence is asymmetric, meaning \(D_{KL}(P||Q) \neq D_{KL}(Q||P)\).

\subsubsection{Making feature maps even more Gaussian}\label{sec:zscore}

For some feature maps, such as logistic regression, the output $y$ represents a probability and lies in the range $[0,1]$, which makes the data unsuitable for fitting a Gaussian mixture model. The data can be transformed using the z-score to make the output from such feature maps more Gaussian. 

The z-score is calculated from the standard normal distribution's quantile function, which is the inverse cumulative distribution function of the normal distribution $\phi^{-1}$ \cite{mccullagh_generalized_2019} (not to be confused with the feature map $\Phi$ used in \cref{sec:EEG_feature_maps}). 
We have probabilities $\Bp$ in the range [0,1], and the z-scores corresponding to these probabilities are given by
\begin{equation*}
 \Bz = \phi^{-1}(\Bp).
\end{equation*}
However, the function is not defined for $p=0$ or $p=1$ because the resulting z-scores would be negative or positive infinity, respectively. This transformation provides an approximation of a Gaussian distribution, but the exact shape of the distribution depends on the predicted probabilities. If these probabilities are mainly close to 0 or 1, the z-scores might take very large or very small values, making the resultant distribution potentially deviate from a standard Gaussian profile. An example of using z-score to make a feature map more Gaussian is found in \cref{fig:zscore}.

For feature maps based on finding a separating hyperplane between the classes, like SVM, the distance to the hyperplane can be used as output $y$ from the feature map instead of the binary classification.

\subsubsection{Gaussian mixture models vs. ensemble learning}
Ensemble learning is a well-known approach for combining the output from different feature maps to improve classification accuracy. One ensemble method is to use a majority vote, and another is to use a weighted sum of the feature maps, but more methods exist \cite{geron_ensemble_2019}. However, in this paper, the combined multi-dimensional Gaussian mixture model is used to predict future epochs for automatic control of stimuli selection in \cref{sec:stimuli_control} and as a basis for transfer learning in \cref{sec:transfer_learning}.

\subsection{Materials and Methods}

\subsubsection{Dataset\label{sec:dataset}}

The Brain Invaders 2013 dataset from the GIPSA-lab, \cite{congedo:11}, was used for empirical analysis. The MOABB Python package was used to access the data \cite{jayaram:18}. The dataset consists of EEG data with 16 channels, 512~Hz, from 24 subjects, whereas the first seven subjects have eight sessions per subject, and the rest performed a single session. Each session consists of several trials with the ratio of target:non-target stimuli as 1:4. The number of trials per session varies due to the experimental setup. The EEG data is recorded in 16 channels, 512~Hz.

The data from the first subject was processed with several feature maps, presented in \cref{sec:EEG_feature_maps}, and Gaussian mixture models were created to estimate the distributions of these feature maps. In general, sessions~1 and 2 were used as training data for the feature maps, session~3 and 4 as validation data for hyperparameter selection, and session~5 as test data. \cref{fig:pdfs} shows the feature map output distribution when session 5 is used as test data.

\subsection{Calculation}

There are many Python toolboxes for EEG epoching, artifact removal, decimation, filtering, etc. MNE\footnote{\url{https://mne.tools}} is a Python toolbox for EEG data processing, and MOABB\footnote{\url{https://https://moabb.neurotechx.com/}} is a toolbox for accessing publicly available data. Some Python packages for calculating the feature maps and GMM estimations are: NumPy\footnote{\url{https://numpy.org/}} for basic math operations, scikit-learn\footnote{\url{https://scikit-learn.org/}} for machine learning-based feature maps and Gaussian mixture model estimation, pyRiemann\footnote{\url{https://pyriemann.readthedocs.io/}} for Riemannian-based feature maps, and SciPy\footnote{\url{https://scipy.org/}} for GMMs. 
These open-source tools support Linux, Windows, and macOS operating systems and run on any decent laptop.

There are many different approaches to EEG preprocessing when it comes to training classifiers for EEG data. We have chosen to downsample the input rate by a factor of 10 from 512~Hz to 51.2~Hz, then apply a fourth-order Butterworth zero-phase band pass filter in the range 0.5~Hz to 20~Hz. Then, we normalize each EEG channel by subtracting the mean and dividing it by the standard deviation. We have not applied any manual or algorithmic artifact removal. After these preprocessing steps, we cut the time series data into epochs, picking the data from stimuli onset to 900~ms after stimuli onset.

For each of the feature maps, there are several hyperparameters to tweak. We have made hyperparameter tuning using validation data for some hyperparameters, such as the C parameter for SVM and LR, but have not done an exhaustive search for all hyperparameters' optimal values for this dataset since the results are sufficient for the purpose of this article without such exhaustive search.

The Kullback-Leibler divergence score is estimated through Monte Carlo sampling with 10000 sample points.

The scripts we provide use around 1~GB of RAM while running, and the complete Brain Invaders 2013 dataset is 27~GB on disk when unpacked. For reproducibility, the Python source code for all results in this paper is published at \url{https://bci.lu.se/automatic}

\subsection{Results\label{sec:Results1}}
\begin{figure}[h!]
\begin{center}
\includegraphics[width=300pt]{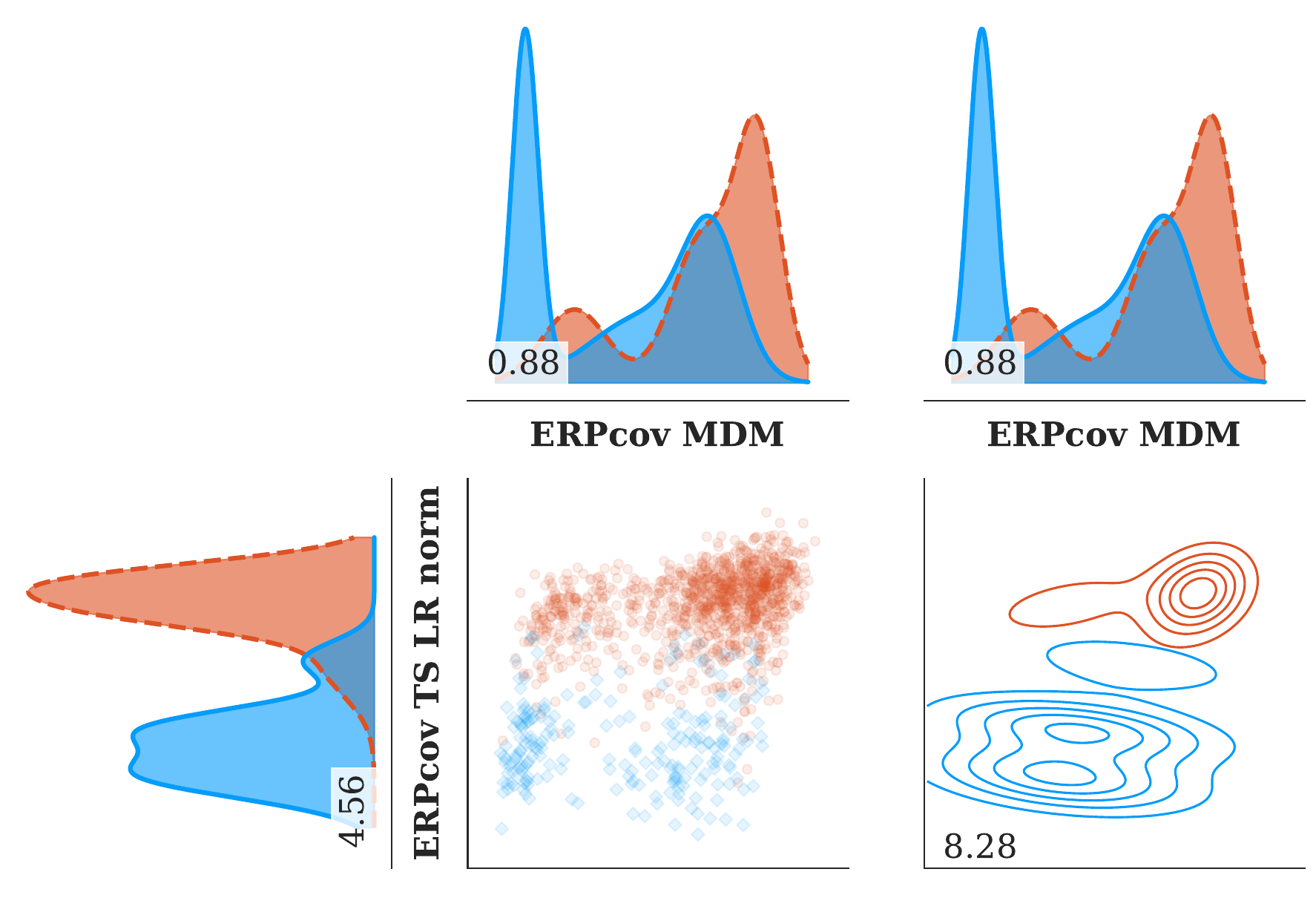}
\caption{Mapping EEG data to real numbers. The scatter plot in the middle shows the data from all 1332 epochs after transformation with two selected feature maps, with blue for target and red for non-target data. The \textbf{ERPcov MDM} feature is used on the x-axis and \textbf{ERPcov TS LR norm} on the y-axis. 
To the left is the fitted 1D GMMs showing an estimation of the distribution of the \textbf{ERPcov TS LR norm} feature map values, and the top shows the \textbf{ERPcov MDM} feature map GMMs. To the right of the scatter plot is the topological representation of the 2D Gaussian mixture model, approximating the data distribution in the scatter plot using a third-order GMM. The numbers in the PDF plots indicate the Kullback-Leibler divergence score for the PDFs and can be interpreted as "the separability of target vs. non-target", as described in \cref{sec:scoring}. The scatter and 2D PDF plots in this figure can be seen in the lower right corner of \cref{fig:pdfs}, and the 1D PDFs are found at the last two rows and columns in \cref{fig:pdfs}.}
\label{fig:eeg_to_gmm}
\end{center}
\end{figure}

The result of mapping EEG data to real numbers is illustrated in \cref{fig:eeg_to_gmm}. The feature maps are trained with training data, hyperparameters are tuned for the feature map with validation data, and the test data after transformation with the feature map is what is shown in the scatter plot in the middle of the figure. The 1D and 2D GMMs are fitted to the transformed test data. See \cref{sec:dataset} for specifics on the dataset. The numbers in the figure show the KL divergence score and indicate the separability between target and non-target GMMs. 

\begin{figure}[h!]
\begin{center}
\includegraphics[width=300pt]{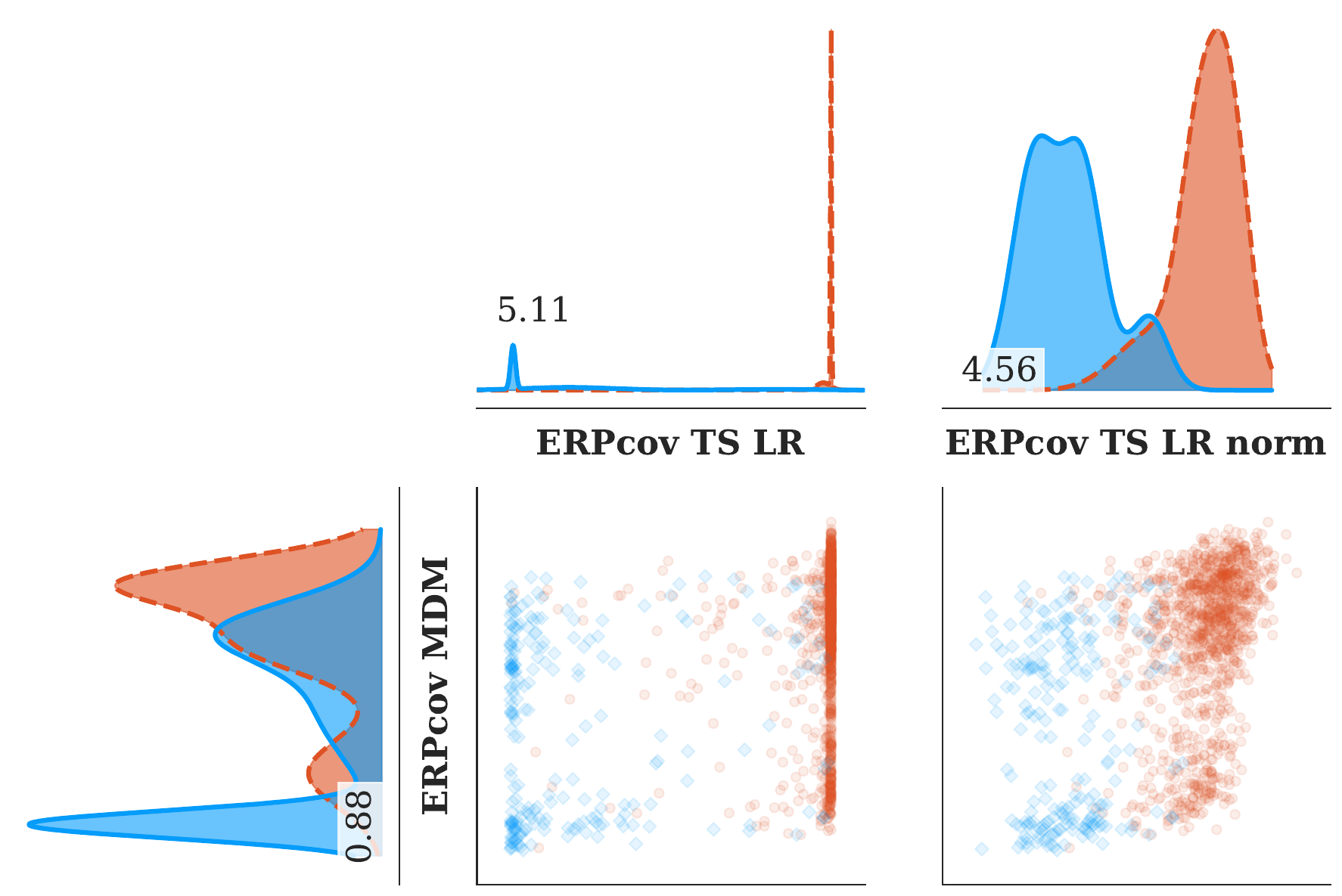}
\caption{Z-score transform to make a distribution more Gaussian. The GMMs to the left with a KL divergence score of 0.88 comes from the \textbf{ERPcov MDM} feature map. The top left GMMs with a KL divergence score of 5.11 comes from the \textbf{ERPcov TS LR} feature map. The top right GMMs with a KL divergence score of 4.56 comes from the \textbf{ERPcov TS LR norm} feature map, which is the z-score transformed data from \textbf{ERPcov TS LR}. In the left scatter plot, the feature map \textbf{ERPcov TS LR} on the x-axis has values representing probabilities $x \in [0,1]$, especially evident when looking at the red non-target values at the right edge of the middle plot.
}
\label{fig:zscore}
\end{center}
\end{figure}

\cref{fig:zscore} shows the results from a z-score transform. 
For the combination of EEG data and \textbf{ERPcov TS LR} feature map that we have, the margin of the closest value to zero is $10^{-6}$, and to one $10^{-9}$. This means that the z-score transform is valid. 
The left scatter plot of \cref{fig:zscore} shows the data before the z-transform, and the right scatter plot the data after. The numbers in the plot show the KL divergence score, which states how well separated the GMMs are, not how well the GMMs are fitted to the data.

\begin{figure}[h!]
\begin{center}
\includegraphics[width=300pt]{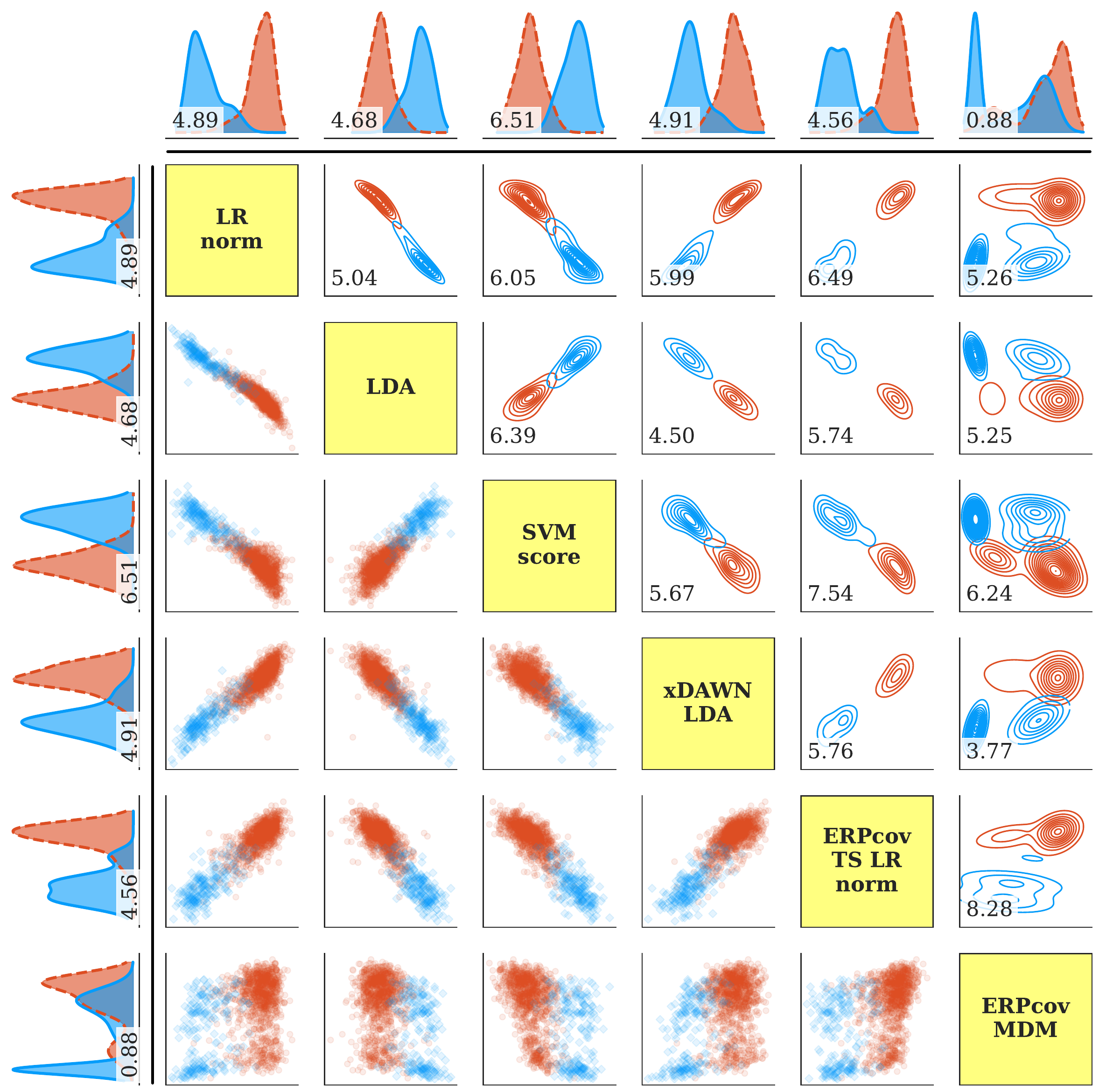}
\caption{Data distributions when mapping the multi-channel EEG epoch data to real-valued outputs using different feature maps, with two-dimensional covariance plots and Gaussian mixture models (GMMs) for target and non-target data. The dataset described in \cref{sec:dataset} is used.
\emph{Diagonal}: The name of the feature mapping. \emph{Top row} and \emph{leftmost column}: 1D GMMs for corresponding feature map in the diagonal. \emph{Lower left triangle}: two-dimensional covariance plots of data from the feature maps in the corresponding row/column. \emph{Upper right triangle}: 2D GMMs for the corresponding row/col feature maps, visualized with level curves from the 2D GMMs.
}
\label{fig:pdfs}
\end{center}
\end{figure}

\cref{fig:pdfs} summarizes the results of generating GMMs from EEG data. The diagonal shows the name of the feature map that is used to map the EEG data to real numbers. The 1D GMMs at the top and to the left of the plot are estimated from the data with the feature map indicated in the diagonal. The lower left triangle of the plot shows scatter plots of data when two feature maps are combined, and the upper right triangle of the plot shows the corresponding topological representations of 2D GMMs. The number in the plots at the upper triangle shows the KL divergence for the target and non-target GMMs for that feature map combination. The higher the KL divergence score, the more separated the GMMs are. If the feature maps were to be heavily dependent, their two-dimensional scatter plots would resemble a thin curve. Looking at the \textbf{ERPcov MDM} feature map, it is obvious that the 2D GMMs better separate the target and non-target distributions. However, it is not as obvious in, for example, the \textbf{LR norm} feature map where the one-dimensional GMMs are relatively well separated. 

\section{Automatic control of stimuli using Gaussian mixture models}\label{sec:stimuli_control}
This section describes the theory, materials and methods, calculations, and results for the automatic selection of the next stimuli based on the GMMs.

\subsection{Theory}

\subsubsection{Bayes' theorem}

The formula for Bayes' theorem, also known as Bayes' rule or Bayes' law, is 
\begin{equation}
    p(x|y)=\frac{p(y|x)p(x)}{p(y)},
    \label{eq:bayes_theorem}
\end{equation}
where the \emph{posterior} $p(x|y)$ is the conditioned probability of $x$ given $y$. The \emph{likelihood} $p(y|x)$ is the conditioned probability of $y$ given $x$, and this is what a maximum likelihood estimator seeks to optimize. The \emph{prior} $p(x)$ describes the prior knowledge of the random variable $x$ which is often an assumed distribution rather than a known distribution. The \emph{evidence} $p(y)$ is a normalization factor, defined as the integration for continuous probabilities, or sum for discrete probabilities, of the numerator over the random variable $x$. See \cite{faisal_probability_2020} for a good introduction to probability and distributions, including a part about Bayes theorem.

\subsection{Materials and Methods}

The paper by \textcite{tufvesson:23} introduced a novel stimulus selection algorithm \emph{Target expectation maximization}. Here, we extend this algorithm to use Gaussian mixture models in multiple dimensions. The main difference between one-dimensional GMMs and multi-dimensional GMMs is that scalars in the 1D case become vectors, which makes the description of the theory below similar to the corresponding sections in the original paper. The multi-dimensional data also allows for covariance analysis between feature maps in a way that is not possible in the 1D case.

\subsubsection{Stochastic stimulus-response model} \label{sec:stimulus_response_model}
This subsection is an extended version of the one in \cite{tufvesson:23}, here rewritten for multi-dimensional features.

We designate the series of $T$ consecutive stimuli as
\begin{equation}
\Bu=\begin{bmatrix}u_1&\hdots&u_T\end{bmatrix}^\top,
\end{equation}
where every stimulus is represented as an integer, signifying one among $C$ distinct and mutually exclusive categories, i.e.
\begin{equation}
u_t\in \{1,\hdots,C\} \triangleq U,\ t=1,\hdots,T.
\end{equation}

We will, in this paper, assume that an observed evoked brain response can be represented as a feature vector 
$\By_t \in \mathbb{R}^n$. Here, $n$ denotes the number of real-numbered features. These features are then organized into a matrix $\BY \in \mathbb{R}^{T,n}$ as

\begin{equation}
\BY=\begin{bmatrix}\By_1&\hdots&\By_T\end{bmatrix}^\top.
\end{equation}

We will also assume that events are independent, such that $\By_t$ is influenced by $u_\tau$ only when $t=\tau$.
We categorize stimuli such that one category, $x \in U$, is labeled as the target. The other $C-1$ categories are non-targets. We aim to determine the hidden variable $x$ using the input-output data $(\Bu,\BY)$.
For responses, $\By_t$ follows a non-target distribution with a probability density function (PDF) $f_0$ when $u_t\neq x$.
If $u_t=x$, it follows a target distribution with PDF $f_1$. For ease of notation, we introduce the compact notation

\begin{equation}
f(\By_t|u_t,x)=
\begin{dcases}
f_0(\By_t)& \text{if}\ u_t\neq x,\\
f_1(\By_t)& \text{if}\ u_t= x.
\end{dcases}
\label{eq:f}
\end{equation}

The likelihood of observing $y_t$ as a response to $u_t$ is denoted as $\L(\By_t| u_t,x)=f(\By_t| u_t,x)$.
As further discussed in \cref{sec:Discussion}, the PDFs $f$ corresponding to calibration sessions with the same subject are static and known. In \cref{sec:EEG_to_GMM}, a few alternatives for calculating these distributions from real EEG data are explained. An approach to transfer learning by gradually personalizing the PDFs  $f_0$ and $f_1$ during the current session is presented in \cref{sec:transfer_learning}.
We express the likelihood of $\BY$ conditioned on the underlying stimuli $\Bu$ and the target category, $x$, being $k$

\begin{equation}
\L_k\triangleq
\L(\BY| \Bu,x=k)=\prod_{t=1}^T f(\By_t| u_t,k).
\label{eq:Lk}
\end{equation}

The maximum likelihood target estimator becomes

\begin{equation}
\hat{x}_{ML}=\argmax_{k\in \{1,\hdots,C\}} \L_k.
\label{eq:x_hat_ML}
\end{equation}

We use Bayes' theorem to compute the probability that a candidate $k$ is the target conditioned on the data, the assumed to be known distributions $f_0$ and $f_1$, and an a priori assumption of equally likely target classes

\begin{equation}
p_k|\Bu,\BY \triangleq 
P(x=k|\Bu,\BY)=\dfrac{\L_k}{\sum_{i = 1}^C\L_i}.
\label{eq:pk}
\end{equation}

At time $t=T$ we can assume to have access to the prior probability for every class $\Bp_{T} = [ p_{1,T}\, \ldots \,p_{C,T}]^\top$, choose the next stimulus $u_{T+1}$, and observe the resulting response $\By_{T+1}$. For clarity, we write $p_{k, T+1}$ as $p^+_k$, $\By_{t+1}$ as $\By$ and $u_{t+1}$ as $u$. The probability of $k$ being the target $p_k$ can then be updated as $p^+_k$:

\begin{equation}
\begin{split}
p^+_k(\By|u,\Bp)&= P(x=k|u,\By,\Bp) \\[4pt]
&=\dfrac{f(\By|u,k) \L_k}{\sum_{i = 1}^C f(\By|u,i) \L_i} = 
\dfrac{f(\By|u,k) p_k}{\sum_{i = 1}^C f(\By|u,i) p_i}.
\label{eq:update}
\end{split}
\end{equation}

We use the vector $\Bp$ to represent probabilities and update the target probability after receiving a new stimulus-response pair. We will now study how we can select the next stimulus $u$ to achieve a more reliable classification from a predetermined number of epochs or how we can minimize the number of epochs needed to achieve a predetermined accuracy.

\subsubsection{Naive candidates for stimuli selection}

When deciding the sequence of stimuli, these are some of the well-known algorithms:

\begin{itemize}
\item \emph{Oracle:}\ Cheat by showing the target stimuli. Always.

\item \emph{Favorite:}\ Choose the $u$ we believe is the most likely target, the one currently with the highest $p$.

\item \emph{Thompson sampling:}\ Assign $u$ randomly, stratified by belief, so that $P(u=i)=p_i$.

\item \emph{Round robin:}\ Given the previous input $u_t$, the next input is chosen as $u_{t+1}=\mod(u_t,C)+1$.

\item \emph{Random:}\ The next input $u_{t+1}$ is drawn from a discrete uniform distribution over $U$.
\end{itemize}

In \cref{sec:Results2} these naive alternatives are compared to the one extended in this paper:
\begin{itemize}
\item \emph{Target expectation maximization (TEM):}\ Choose $u$ to maximize the next true target probability $p^+_x$. This algorithm was introduced in \cite{tufvesson:23}.
\end{itemize}

\subsubsection{Target expectation maximization (TEM)}

When choosing a stimulus candidate $u$, we don't yet know the resulting response $\By$. However, we understand that 
$\By$ will match the target distribution with a probability $p_u$. It will follow the non-target distribution with a probability $1-p_u$. Presenting the stimulus $u$, the likelihood of the resulting response $\By$ is 

\begin{equation}
\L(\By|u)=p_u f(\By|u,x=u) + (1-p_u)f(\By|u,x\neq u). 
\label{eq:LH_}
\end{equation}

The likelihood $\L(\By| u)$ integrates to unity over $\By$, meaning $\By$ will follow a distribution with the PDF

\begin{equation}
g(\By|u)=\L(\By|u).    
\end{equation}

By treating $p^+_k$ of \cref{eq:update} as an ordinary function of the stochastic variable $\By \in \mathbb{R}^n$, we can compute the expectation

\begin{equation}
p^*_k(u)\triangleq\mathbb{E}\ p^+_k(\By|u) = \int_{\mathbb{R}^n} p^+_k(\By|u)g(\By|u)\ d\By,
\end{equation}
which is an ordinary function of $u$.

A natural candidate for the next stimuli is the one maximizing the expectation of the updated true target probability $p_x$. Since we do not know which candidate is the target, we will rely on our prior belief $\Bp$ and choose

\begin{equation}
u^* = \argmax_{u\in \{1,\hdots,C\}} \sum_{k = 1}^C p_k
p^*_k(u).
\end{equation}

\subsubsection{Dataset}
We have used the same dataset for stimuli selection as described in \cref{sec:dataset}. The GMMs were generated using data from subject~\#1. Data from sessions 1 and 2 were used for training data, sessions 3 and 4 for validation data, and session 5 for test data. The used feature maps were \textbf{ERPcov MDM} and \textbf{ERPcov TS LR norm}.

\subsection{Calculation}

Julia was chosen as the implementation language in the paper \cite{tufvesson:23} where the TEM algorithm was introduced. Unfortunately, the Julia package used for GMM calculations (Distributions.jl) does not yet support multivariate GMMs, so we chose to reimplement the same algorithms in Python instead. 

\subsubsection{Runtime and download}
The TEM algorithm has a low computational cost in the 1D case, which allows for real-time decisions about stimuli selection on embedded systems. We ran 2048 Monte Carlo simulations for each algorithm using different random seeds, ensuring varied samples from the probability density functions. The original code in \cite{tufvesson:23}, written in Julia, runs in under a minute. The same algorithm implemented in Python takes ten times longer, mainly due to Python's sanity checks around calls to external code (the integration algorithms in SciPy are using legacy Fortran code\footnote{\url{https://docs.scipy.org/doc/scipy/reference/generated/scipy.integrate.dblquad.html}}), which then also makes frequent calls back to Python code for the function to be integrated. When switching from 1D integrations to the 2D TEM algorithm, the runtime is 3.5 hours in Python (20x compared to the 1D version), meaning additional engineering and optimization needs to be done to reduce the runtime for it to be a viable method in a real-time BCI. A starting point would be to investigate very small floating-point numbers, known as subnormal numbers\footnote{\url{https://en.wikipedia.org/wiki/Subnormal_number}}, as modern CPUs struggle with calculations involving these. The Python source code for reproducing the results in this paper can be found at \url{bci.lu.se/automatic}

\subsection{Results\label{sec:Results2}}

In evaluating algorithm performance, we can track how many stimuli it takes to achieve 95\

\begin{figure}
\begin{center}
\begin{subfigure}[t]{0.48\linewidth}
\includegraphics[width=\textwidth]{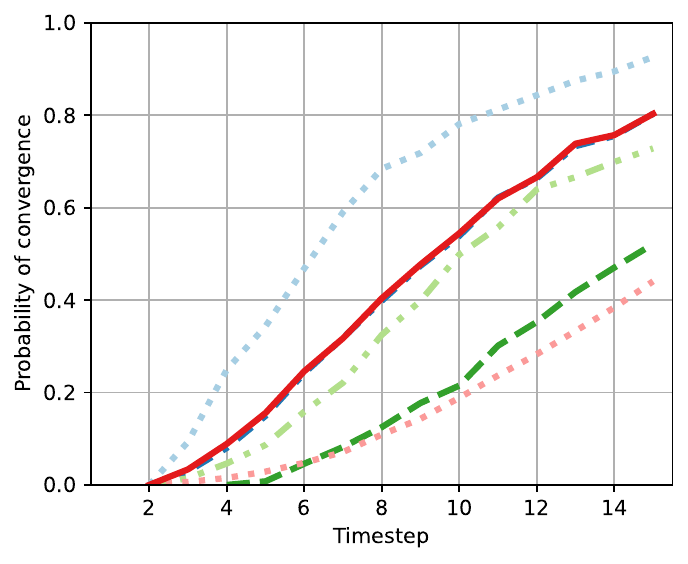} 
    \caption{Using 1D GMMs with \textbf{ERPcov MDM} having a KL divergence score of 0.88 as feature map. Note that the timescale is different in this plot compared to the others.}
    \label{subfig:cdf_1DGMM_ERPcov_MDM}
\end{subfigure}
\hfill
\begin{subfigure}[t]{0.48\linewidth}
\includegraphics[width=\textwidth]{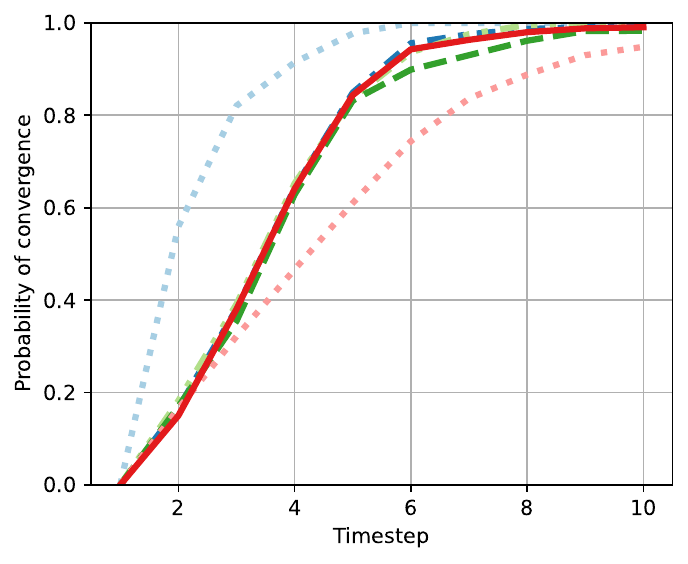} 
    \caption{Using 1D GMMs with \textbf{ERPcov TS LR norm} having a KL divergence score of 4.56 as feature map.}
    \label{subfig:cdf_1DGMM_ERPcov_TS_LR_norm}
\end{subfigure}

\begin{subfigure}[t]{0.48\linewidth}
\includegraphics[width=\textwidth]{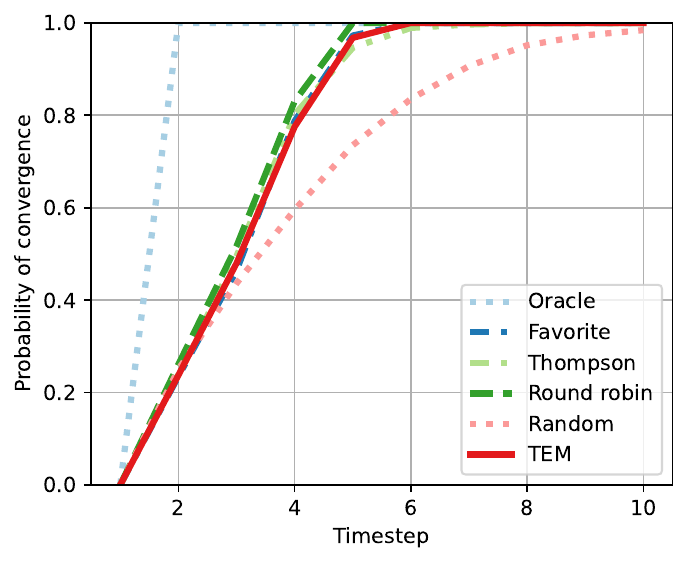} 
    \caption{Using 2D GMMs with \textbf{ERPcov TS LR norm} and \textbf{ERPcov MDM} having a combined KL divergence score of 8.28 as feature maps.}
    \label{subfig:cdf_2DGMM}
\end{subfigure}

\caption{Probability that a session correctly states one out of four colors with a certainty of 0.95 as a function of the number of timesteps, using six different stimuli selection algorithms. The plots show the results of 2048 Monte Carlo simulations. The TEM algorithm was introduced for the one-dimensional GMM case in \cite{tufvesson:23}.} 
\label{fig:cdf}
\end{center}
\end{figure}

\section{Transfer learning using Gaussian mixture models}\label{sec:transfer_learning}
This section describes the theory, materials and methods, calculations, and results for adapting a priori knowledge of GMMs to fit the data from a new session with unknown labels. In other words, how a reactive BCI based on GMMs from feature maps can be made calibration-free. 

\subsection{Theory}

Before a BCI can be used, it must recognize the user's brain signals. Brain signals differ between sessions and persons due to different mental states, other ongoing brain activity, surrounding noise, slightly different electrode placement, etc. This means that a BCI must be fine-tuned for a new session every time it is used. During the so-called calibration of BCIs, labeled data is collected, and the underlying machine learning algorithms of the BCI are trained on that data \cite{nam_transfer_2018}. 

The calibration phase of a BCI takes a long time and is tiresome for the user. Thus, it is desirable to reduce the calibration time of BCIs. One approach to do this is to use transfer learning to utilize data from previous sessions or users when training the machine learning algorithms \cite{lotte_signal_2015}. In the language of transfer learning, the previous data is called \emph{source data}, and the new data is called \emph{target data}. In this paper, target already refers to target or non-target data in the sense of attended stimuli. Therefore, we will use the terms \emph{source data} and \emph{destination data} to avoid notation overload. There are several approaches to transfer learning, but they can generally be done either on the data, \emph{domain adaptation}, or on the algorithms, \emph{rule adaptation} \cite{nam_transfer_2018}. One method for domain adaptation for EEG data is the Riemannian Procrustes analysis algorithm \cite{rodrigues_riemannian_2019}. In this paper, a variant of rule adaptation is made on the distributions $f_0$ and $f_1$. The GMMs from source data are denoted $f^s_0$ and $f^s_1$, and their parameters are updated with an online variant of the expectation-maximization algorithm to fit the destination data. There are many variants of the online update algorithms, sometimes called sequential expectation-maximization algorithms or maximum a priori parameter estimation algorithms, but common to all is that parameters of the GMM are updated as a weighted sum of the parameters of the source GMM and estimated parameters from the destination data \cite{awwad_shiekh_hasan_sequential_2009, reynolds_gaussian_2015, zivkovic_recursive_2004}.

\subsection{Materials and Methods}

Assuming that the GMMs from the source domain data, $f^s_0$ and $f^s_1$, are somewhat similar to the unknown GMMs from the destination domain data, $f^d_0$ and $f^d_1$, the source domain GMMs can be used as a priori information for transfer learning. \cref{subfig:transferlearning_iteration_0} shows such a situation where the GMMs for source and destination data are similar but not the same and where this approach for transfer learning is suitable. In overview, new destination data $\By$ arrives consecutively in an online BCI setting, and the GMMs are successively updated to fit the destination data.

The labels of the destination data from an ongoing BCI trial are generally unknown, and the first step to be able to update the GMMs is to predict the labels of the destination data based on the current GMMs. In our case, the ratio of target and non-target stimuli can be known, since we decide the stimuli ourselves. Only one stimuli class is the target, and for instance choosing a uniform distribution for stimuli will result in one single known target vs non-target ratio, regardless which class is the target. Thus, the distribution of data label predictions should follow the stimuli distribution. We reclassify those samples with the highest probability of belonging to the other class if the predicted labels don't follow the expected distribution. However, if the label is known, no prediction is needed. In either case, the tuple $(\By,l)$ connects a label, $l$, to the destination data, $\By$. For destination data with unknown labels, all labels are re-predicted every time new destination data arrives to update the predictions according to the current GMM estimates. 

Once all data have labels, the parameters for GMM $f_i$ are updated using all available destination data labeled $l=i$. We denote the subset of the dataset with labels $l=i$ as $\BX$, and the next step is predicting which Gaussian $j$ in the Gaussian mixture model the data $\BX$ belongs to. We denote the data $\BX$ belonging to Gaussian $j$ as $\BX_j$. The parameters for the $j^\text{th}$ Gaussian in GMM $i$, which are mixing coefficient($\pi_j$), mean ($\Bmu_j$), and covariance matrix ($\BSigma_j$), can be updated as:
\begin{align*}
    \pi_j^+ &= (1-\alpha)\pi^s_j + \alpha\hat{\pi}_j \\
    \Bmu_j^+ &= (1-\alpha)\Bmu^s_j + \alpha\hat{\Bmu}_j \\
    \BSigma_j^+ &= (1-\alpha)\BSigma^s_j + \alpha\hat{\BSigma}_j,
\end{align*}

where any variable with a hat ($\;\hat{}\;$) means estimated parameters from destination data $\BX_j$, superscript $s$ means parameters from source domain GMM (a priori knowledge), and superscript $^+$ the new parameters. $\alpha$ is the learning rate for the update and is here defined as $\alpha = 1-(1+bn)e^{-bn}$, where $e$ is Euler's number used for the exponential function, the coefficient $b$ is chosen as 0.03 for this dataset and $n$ is the number of samples in $\BX_j$.
The estimated parameters are calculated as:
\begin{align*}
\hat{\pi}_j &= \frac{n}{\text{count}(\BX)} \\
\hat{\Bmu}_j &= \frac{1}{n} \sum^n_{k=1} \Bx^k_j \\
\hat{\BSigma}_j &=
\begin{dcases}
\text{Identity matrix}& \text{if}\ n \leq 1,\\
\frac{1}{n} \sum^{n}_{k=1}(\Bx^k_j - \hat{\Bmu}_j)(\Bx^k_j - \hat{\Bmu}_j)^{\top}, & \text{if}\ n > 1,
\end{dcases}
\end{align*}

where $\Bx^k_j$ means the k-th row in the matrix ${\BX}_j$.

\subsubsection{Dataset}\label{sec:dataset_transfer_learning}
We have used the same dataset as described in \cref{sec:dataset}. For intra-subject transfer learning, data from subject \#1 sessions 1 and 2 are used for training the feature maps, and data from sessions 3 and 4 for hyperparameter selection. Data from sessions 5 and 6 are used as test data where the data from session 5 is considered source data and data from session 6 is destination data. See \cref{fig:transfer_learning_results} for results.

For cross-subject transfer learning, data from subjects~\#1 and \#7 are used. The data from sessions 1 and 2 are used for training the feature maps, data from sessions 3 and 4 for hyperparameter selection, and data from session 5 are used as test data. See \cref{fig:transfer_learning_2_subjects_results} for results.

\subsection{Calculation}

The transfer learning method is implemented using Python and the NumPy\footnote{\url{https://numpy.org}}, scikit-learn\footnote{\url{https://scikit-learn.org}}, and SciPy\footnote{\url{https://scipy.org}} Python packages. Matplotlib\footnote{\url{https://matplotlib.org}} is used for plotting, and ffmpeg\footnote{\url{https://ffmpeg.org/}} for generating the video. The algorithm itself can easily be run in an embedded environment, and the main calculation time is spent doing plotting. The source code for reproducing the results is available at \url{https://bci.lu.se/automatic}

\subsection{Results\label{sec:Results3}}

\begin{figure}[h!]
    \centering
    \begin{subfigure}[t]{0.48\linewidth}
        \includegraphics[width=\textwidth]{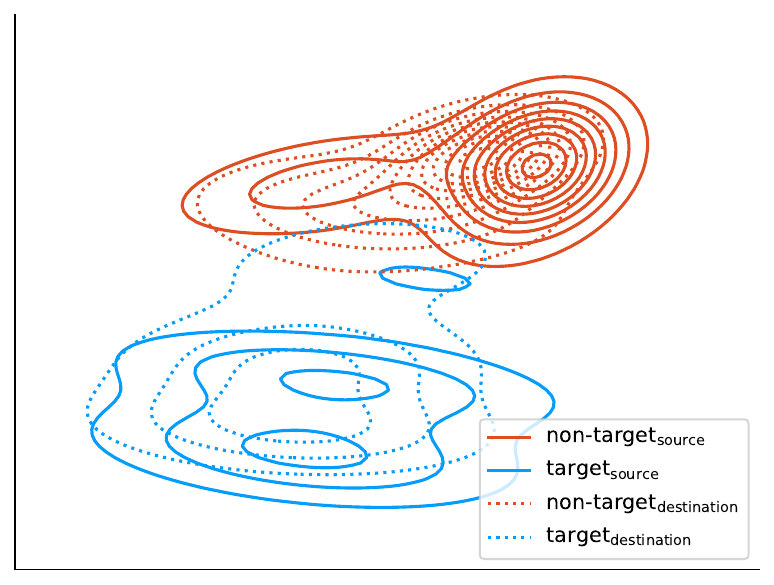}
        \caption{Distributions before transfer learning.}
        \label{subfig:transferlearning_iteration_0}
    \end{subfigure}
    \hfill
    \begin{subfigure}[t]{0.48\linewidth}
        \includegraphics[width=\textwidth]{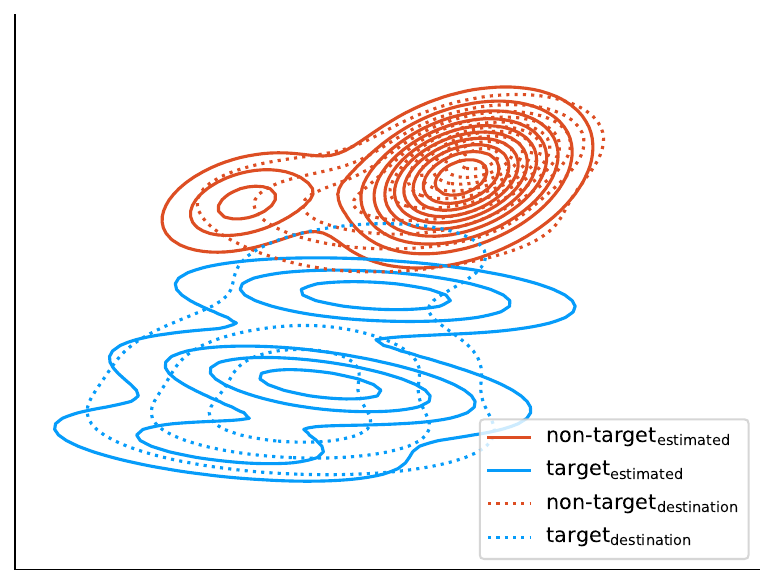} 
        \caption{Snapshot of distributions during transfer learning halfway into a session.}
        \label{subfig:transferlearning_iteration_600}
    \end{subfigure}
    \caption{Transfer learning level plots for the intra-subject distributions of target (blue) and non-target (red) data. Dotted lines are destination distributions and solid lines are in (a) source and (b) estimated GMMs during transfer learning. A video of iterations of this plot during transfer learning can be found here: \url{https://bci.lu.se/automatic}. The source GMMs are successfully moved towards the destination GMMs during the transfer learning.}
    \label{fig:transfer_learning_results}
\end{figure}

Transfer learning using Gaussian mixture models is a viable approach toward calibration-free reactive brain computer interfaces. \cref{fig:transfer_learning_results} shows two frames from the video\footnote{\url{https://bci.lu.se/automatic}} showing transfer learning using the intra-subject data described in \cref{sec:dataset_transfer_learning}. 

\begin{figure}[h!]
    \centering
    \includegraphics[width=300pt]{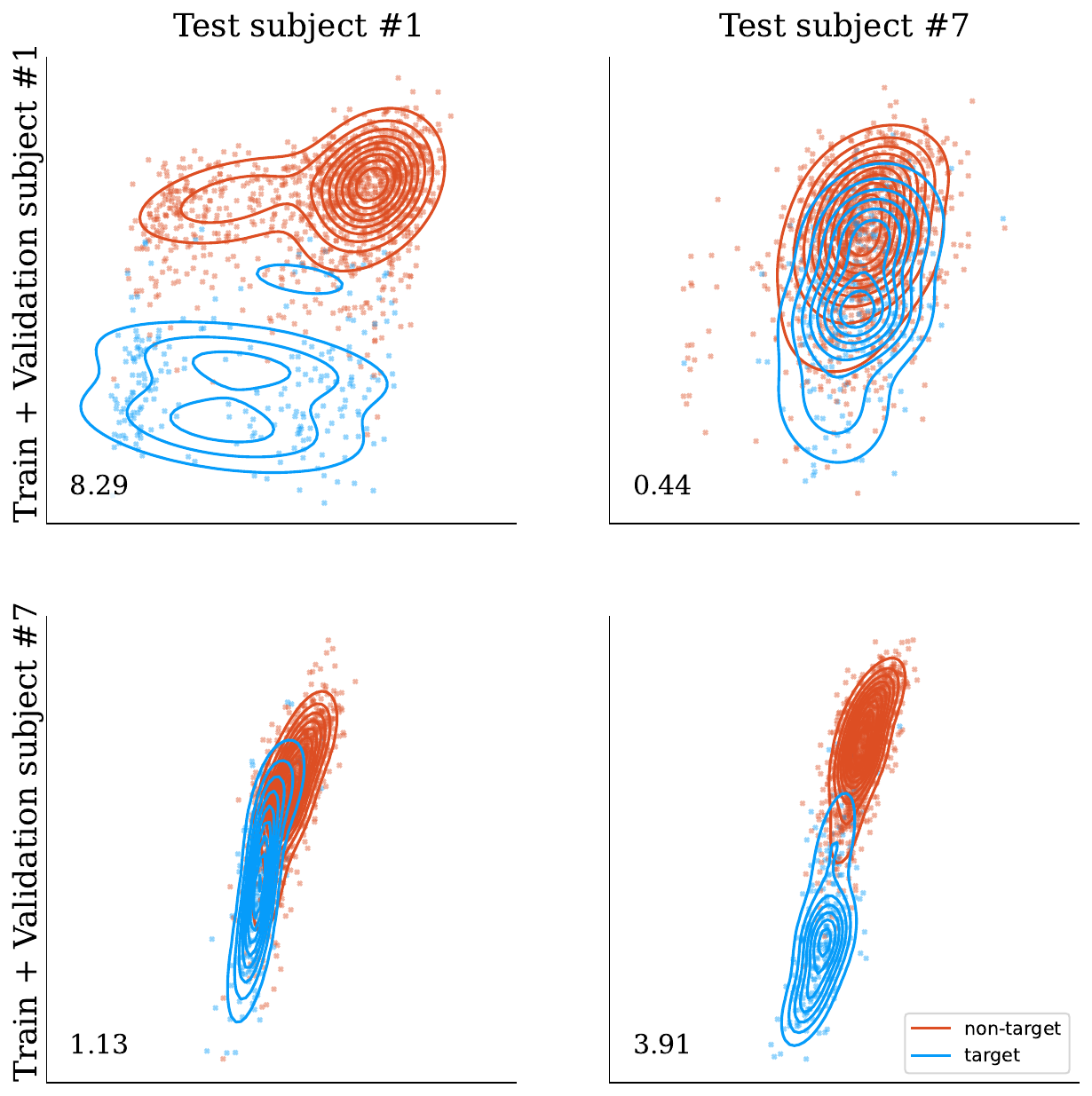}
    \caption{
    Transfer learning plots showing data from feature maps \textbf{ERPcov TS LR norm} and \textbf{ERPcov MDM} trained on sessions 1 and 2, and validated on sessions 3 and 4 for subjects~\#1 and \#7 as indicated in the rows. Then, session 5 from either subject~\#1 or subject~\#7 is used to test the data as indicated in the columns. The number shows the KL divergence score for the GMMs. The intra-subject GMMs on the diagonal are well separated but the cross-subject GMMs on the off-diagonal are not well separated and transfer learning in these cases is not a viable approach, see the discussion below.}
    \label{fig:transfer_learning_2_subjects_results}
\end{figure}

\cref{fig:transfer_learning_2_subjects_results} shows GMMs for two different users where the training and test data is intra- and cross-subject, as indicated in the figure. The plots on the diagonal show intra-subject GMMs, meaning that data from the same user but different sessions are used for training, validation, and testing. The off-diagonal plots show cross-subject GMMs, meaning that data from one subject is used for training and validation, and data from another subject is used for testing. The intra-subject GMMs have a higher KL divergence score than the cross-subject GMMs. 

\section{Discussion\label{sec:Discussion}}

\cref{sec:stimuli_control} of this paper extends the paper by \textcite{tufvesson:23}, embracing the idea of using Gaussian mixture models (GMMs) for stimuli selection but expanding them into multi-dimensional GMMs. This paper focuses on 2D GMMs, but there is no hard limit to the GMMs' dimensions. However, higher-order GMMs have a higher CPU cost for calculations. As stated above, the theoretical results and reasoning are similar for 2D as for higher dimensions, and for the sake of printability on paper, the results are illustrated with 2D GMMs. 

\cref{fig:eeg_to_gmm} shows the output from two feature maps and the corresponding 1D and 2D GMMs for the distribution of this output. We use three Gaussians in the GMMs and fit the GMMs to the data using the expectation-maximization (EM) algorithm. For the purpose of this paper, the evaluation of how well a certain GMM fits the data would not necessarily contribute to the discussion. Since, we explore how GMMs can be used in a reactive BCI and have no need to find the optimal feature map or GMM for this data. However, we use real EEG data to get realistic, but not necessarily optimal, results for the feature maps and fitted GMMs.

Studying the results in \cref{fig:zscore}, it appears that the fitted 1D GMM to the data from the \textbf{ERPcov TS LR} feature map is better than the data from the \textbf{ERPcov TS LR norm} feature map. While it is true that the KL divergence score for \textbf{ERPcov TS LR} is higher than for \textbf{ERPcov TS LR norm}, one must remember that the KL divergence score states how well the target and non-target GMMs are separated, and not how well the GMM is fitted to the data. The motivation for the z-score transformation is that the data, shown in the left scatter plot in \cref{fig:zscore}, is not suited for GMMs, thus the need to make the data more Gaussian with the z-score transform.  

When fitting GMMs to data, the number of Gaussians is a hyperparameter to choose. Throughout this paper, three Gaussians are used for the GMMs. Looking at \cref{fig:pdfs}, it is obvious that fewer Gaussians would be enough for some feature maps (for example, \textbf{LDA} and \textbf{xDawn LDA}) while others might need more Gaussians (for example, \textbf{LR norm} and \textbf{ERPcov MDM}). To reduce the complexity of the paper, we omitted the number of Gaussians as a parameter to tune and kept it the same for all feature maps. It is worth noting that the GMMs are fitted to real data and using more data would give a better fit. In the dataset we have used, the target:non-target data ratio is 1:4, as described in \cref{sec:dataset}, which means that the non-target GMM is fitted to four times more data than the target GMM. This makes the target GMM more sensitive to the initial random state of the estimator than the non-target GMM. This difference in data availability is unavoidable in a reactive BCI setting since, by design, a greater number of stimuli are non-target than target.

The selection of feature maps presented here illustrates that any function that maps EEG data to real numbers could work as a feature map. The best selection of feature maps depends on the experiment, and there is no ``silver bullet'' that works best for all experiments. The scoring of the GMMs provides some guidelines for choosing feature maps. As presented in \cref{sec:scoring}, there are many available metrics to compare two GMMs. We chose to use the Kullback-Leibler (KL) divergence as a metric hoping we could compute it analytically, which works well with two 1D Gaussian distributions, but unfortunately not with multi-dimensional GMMs. The higher the KL score is, the more separated the GMMs are. As discussed above, the GMMs are estimated from data, so the KL score should be seen as a guiding value rather than a strict decision rule. Computing the separability of the feature map output, meaning the true data, would be a better metric. In other words, if two different pairs of GMMs give approximately the same KL score, other aspects of the feature maps should be considered when choosing the feature maps, such as computational time, risk of overfitting to training data, etc.

The Kullback-Leibler scores we print in our figures are directly comparable between 1D GMMs, and between 2D GMMs. However, comparing KL divergence scores between 1D and 2D GMMs can be problematic due to the inherent differences in the dimensionality of the data. Also, note that the scores reflect the relative entropy between the fitted GMMs, which we have chosen to consist of three Gaussians throughout this paper, and not between the true feature maps. This could explain why a 2D feature map sometimes has a lower KL divergence score compared to its 1D counterparts.

In the realm of real-time BCI, the selection of subsequent stimuli plays a pivotal role in determining the probability of accurately classifying the target category. The TEM algorithm was presented using one-dimensional GMMs in the paper by \textcite{tufvesson:23} where it is compared to a few well-known stimuli selection algorithms. This paper extends that paper and uses two-dimensional GMMs to showcase that the TEM algorithm also works with higher-dimensional GMMs. The expected result of using 2D GMMs instead of 1D GMMs is that the convergence rate for predicting the target stimuli would increase significantly since the 2D GMMs can separate the distributions better than 1D GMMs due to the extra data dimension. The first conclusion to draw from \cref{fig:cdf} is that using 2D GMMs works. Thus, we have shown that the TEM algorithm can be extended to multi-dimensional GMMs. The results in \cref{fig:cdf} also show, as expected, that the Oracle algorithm always outperforms all other algorithms. This is expected since the Oracle knows the target stimuli. The results also show that randomly selecting stimuli is always the worst approach, which is also expected because the random approach does not use any gained information from brain reactions to previous stimuli.

When comparing the results from 1D GMMs, \cref{subfig:cdf_1DGMM_ERPcov_TS_LR_norm,subfig:cdf_1DGMM_ERPcov_MDM}, to the result in Fig.~8 in \cite{tufvesson:23}, it is important to note that different feature maps have been used. The 1D GMM distributions from the \textbf{ERPcov MDM} feature map used in \cref{subfig:cdf_1DGMM_ERPcov_MDM} is more similar to the distribution used in \cite{tufvesson:23} than the distribution from the \textbf{ERPcov TS LR norm} feature map used in \cref{subfig:cdf_1DGMM_ERPcov_TS_LR_norm}, see 1D distributions in \cref{fig:pdfs} and Fig.~6 in \cite{tufvesson:23}. Since the underlying GMM distributions for \cref{subfig:cdf_1DGMM_ERPcov_MDM} are similar to the ones in \cite{tufvesson:23}, it is not surprising that the results are similar. In \cref{subfig:cdf_1DGMM_ERPcov_MDM}, we see that the TEM algorithm is as good as the favorite algorithm and better than the Thompson and round-robin algorithms. The results in \cref{subfig:cdf_1DGMM_ERPcov_TS_LR_norm} are significantly better than in \cref{subfig:cdf_1DGMM_ERPcov_MDM}, which results from the underlying target and non-target GMMs being more separated in \cref{subfig:cdf_1DGMM_ERPcov_TS_LR_norm} than \cref{subfig:cdf_1DGMM_ERPcov_MDM}. From this result, we can conclude that the more separated the target and non-target GMMs are, the better the performance, which comes as no surprise. 

In \cref{subfig:cdf_2DGMM} the 2D GMMs from the combined feature map of \textbf{ERPcov MDM} and \textbf{ERPcov TS LR norm} is used, the same feature maps as in \cref{subfig:cdf_1DGMM_ERPcov_TS_LR_norm,subfig:cdf_1DGMM_ERPcov_MDM} respectively. We see that using 2D GMMs results in a faster convergence for identifying the target stimulus than for the 1D cases, which indicates that using higher-dimensional GMMs improves performance. Interestingly, most algorithms in \cref{subfig:cdf_2DGMM} have identified the target stimulus after showing five stimuli. This experiment's target stimulus is a specific-colored silhouette out of four colors. Bayesian statistics and specifically the update of probabilities by Bayes' theorem will change the probability for the stimulus class vs. the probability for "the others". We will need to show at least three different stimuli to be able to separate the probabilities for all our classes. However, when the target and non-target GMMs are well separated, as in the case of our 2D GMMs, we might get lucky and show the target stimuli as the first one, resulting in a probability of 0.25 after showing the first stimulus, as seen at timestep 2 in \cref{subfig:cdf_2DGMM}. The same reasoning is why the Oracle algorithm, which always shows the target stimulus, reaches the probability of 1.0 after the first stimulus in \cref{subfig:cdf_2DGMM} but not in \cref{subfig:cdf_1DGMM_ERPcov_MDM} or \cref{subfig:cdf_1DGMM_ERPcov_TS_LR_norm}. For the cases in \cref{subfig:cdf_1DGMM_ERPcov_TS_LR_norm} and \cref{subfig:cdf_2DGMM}, where we have an excellent separation of the target and non-target feature maps, the round robin stimuli sequence algorithm performs best, not only because it has zero computational cost, but mainly since it will always cycle through all stimuli. The Thompson, TEM and Favorite algorithms may show the same stimulus twice, which is beneficial in the case seen in \cref{subfig:cdf_1DGMM_ERPcov_MDM} where the KL divergence score is low, but is a drawback for the other cases. In a commercial setting, you would probably start reasoning about cost reductions like reducing the number of EEG electrodes or switching to more user-friendly and convenient dry EEG electrodes, leading to less separable target and non-target GMMs, and then choosing a proper stimuli sequence algorithm would make sense again.

In the 2D case, the target stimulus is identified after showing each stimulus once. This results from the target and non-target GMMs being well separated in the 2D case. Noteworthy is that the same performance could be achieved with well-separated 1D GMMs as well, but 2D GMMs are generally better separated than 1D GMMs due to the extra dimension, given that the same feature map is used in both the 1D and 2D cases.

The biggest drawback of using 2D GMMs compared to 1D GMMs is the computational time, which is about 20 times longer for 2D-based algorithms. Such long computational times would make using 2D GMMs in a real-time BCI infeasible. One runtime optimization would be to run multiple 1D GMMs in series (reusing the same EEG data from one epoch) instead of the 2D combined. Thus, it can be concluded that 1D GMMs, which separate the target and non-target data well, are the best to use. If no such 1D GMMs exists, 2D GMMs could be used.

From \cref{subfig:transferlearning_iteration_0}, it is clear that the GMMs differ even between sessions for an individual. Different GMMs between sessions mean that the assumed distribution when selecting the optimal stimuli, as in the TEM algorithm, and the actual (unknown) distribution the data comes from during the operation of the BCI are different. This, in turn, means that the algorithm for optimal selection of stimuli is based on outdated GMMs. To overcome this pitfall, transfer learning can be used to update the assumed (also a priori knowledge) GMMs based on the received data. The result from such an update is seen in \cref{subfig:transferlearning_iteration_600}, and visually it seems like the transfer learning moves the source GMMs to the destination GMMs.

A ready-to-use BCI means, in this context, that a user can start operating the BCI without calibration of the BCI, given that the feature maps are trained already. In practice, this means that the user has gone through a calibration phase in the past that is used for training the feature maps and may, for every new session, start using the BCI immediately. A fully calibration-free BCI means, in this context, a BCI that requires no calibration at all. In practice, that would mean that the used feature maps are trained on data from another subject than the new user. \cref{fig:transfer_learning_2_subjects_results} shows intra- and cross-subject GMMs. The cross-subject shows the case of fully calibration-free BCI. The KL divergence score is low for these cases, indicating that training and validating the feature maps on one subject and then using them on another is unsuitable in this case. This result is not surprising because it is known that a model trained on one subject rarely works for another subject; thus, it is necessary to calibrate BCIs \cite{lotte_signal_2015}. A ready-to-use BCI is, however, possible to create with the methods presented in this paper. As discussed above, GMMs from previous sessions for one subject are used as a priori knowledge for online updates of the GMMs to the new session.

This paper suggests using adaptive updates of the GMMs as a transfer learning method for reactive BCIs. It is a kind of rule adaptation transfer learning since the parameters of the GMMs are moved and not the data in itself. Another approach for rule adaptation would be to update the feature maps, which are trained on the source data, to fit the destination data. For example, moving the hyperplane in a LDA-based feature map. The EEG experimental setup in this paper is inherently low on subject-specific EEG data, and the transfer learning approach needs to work under severely data-constrained settings, which the suggested GMM transfer learning approach does.

Our research as presented in this paper is aiming for a calibration-free brain computer interface, utilizing GMMs mainly for their adaptability when dealing with probability density functions of feature maps. The success in other machine learning fields, like image recognition and natural language processing, may be repeated in the brain computer interface research field when ubiquitous consumer devices provide an adequate amount of training data.

\section{Conclusions} \label{sec:conclusions}

This journal paper explores using automatic control in reactive brain computer interfaces. A grand challenge for current BCI research is establishing efficient methods for reliable online classification of neural activity. This paper uses Gaussian mixture models for automatic stimuli selection in reactive BCIs, which improves the online performance of BCIs. Our approach is a step towards ready-to-use brain computer interfaces with the potential of expanding the boundaries of BCI appliances and research.

\section*{Acknowledgements}

Thanks to Martin Gemborn Nilsson, Kristian Soltesz, and Bo Bernhardsson at Lund University for contributing to the original idea of the TEM algorithm.

\section*{Contributions}

Pex Tufvesson: Conceptualization, Formal Analysis, Investigation, Methodology, Software, Validation, Visualization, Writing – original draft, and Writing – review \& editing. Frida Heskebeck: Conceptualization, Formal Analysis, Investigation, Methodology, Software, Validation, Visualization, Writing – original draft, and Writing – review \& editing.

\section*{Funding}
The authors are members of the ELLIIT Strategic Research Area at Lund University. This work was partially supported by the Wallenberg AI, Autonomous Systems and Software Program (WASP) funded by the Knut and Alice Wallenberg Foundation.

\section*{Conflict of interest statement}
The authors declare that the research was conducted in the absence of any commercial or financial relationships that could be construed as a potential conflict of interest.

\printbibliography

\end{document}